# GUIDELINE FOR THE PRODUCTION OF DIGITAL RIGHTS MANAGEMENT (DRM)

Shannon Kathleen Coates and Hossein Abroshan

School of Computer and Information Sciences, Anglia Ruskin University, Cambridge CB1 1PT, UK

## ABSTRACT

*Multiple news sources over the years have reported on the problematic effects of Digital Rights Management, yet there are no reforms for DRM development, simply removal. The issues are well-known to the public, frequently repeated even when addressed: impact on the software and to the devices that run them. Yet few, if any, have discussed it in recent years, especially with the intent of eliminating the shown issues. This study reviews Digital Rights Management as a general topic, including the various forms it can take, the current laws that affect DRM, and the current public reception and responses. This study describes the different types of DRM in general terms and then lists both positive and negative examples.*

## KEYWORDS

*Digital Rights Management, DRM implementation, Copyright Protection*

## 1. INTRODUCTION

Digital Rights Management is a very commonly used factor in many digital and occasionally physical products, both public and private. It is intended to protect copyright holders from having their work distributed without their consent, with minimal issues for the users who purchase it legitimately. However, in recent years, the public has seen many examples of DRM disrupting their software/hardware. As such, many now take a negative stance towards it, oftentimes automatically.

Surprisingly, there have been no attempts to dissuade this opinion. If DRM appears to cause issues, it is merely removed from the software in question. No attempts to rectify it while still protecting the product or even to prevent similar issues from recurring. Thus, this is the question posed:

"What negative implementations of DRM exist, and what would need to be done to prevent them from reoccurring?"

The following article will be divided into several sections and will aim to describe each of DRM alongside their positive and negative examples. The negative examples will be then sorted into the specific issues they cause, then into a basic set of guidelines of what DRM should not be able to do with the attached software/end device. This would restore public confidence in Digital Rights Management and prevent future incidents while also ensuring it can still function as intended.

## 2. LITERATURE REVIEW

With the use of various methods and tools, and guidelines, successful digital rights management prevents the inappropriate use of the content it is linked to. A DRM system is used to ensure that only authorised and licenced users can access the bought/owned material as defined by the guidelines on their licence [1-3]. This allows the producer to designate their rights and extract the necessary metadata, and the consumer can choose their material and has a variety of possibilities for content use while additionally allowing the content creator to monitor payment data and content usage [1].

These technologies were initially based on completed work from a European Commission sponsored project called Imprimatur. The outcome of which was a business model for the assessment of RMI1 and the digital content distribution [4]. Customers, retailers, copyright holders, service providers, and artists were considered the primary players in the market. They can exchange information on copyright and content using electronic distribution channels [4]. As a consequence, managing and protecting these copyrights is important for all parties involved. There have been multiple studies into protecting these digital rights, even when several parties are involved in the distribution and sub-distribution of these products, some of which may not even be aware of the areas they are be distributed to [5-7].

A study published in 2021 [8] focuses on digital rights management in regards to e-books2, specifically one that assessed "how the removal of DRM systems would change the consumer surplus in the e-book platform". It acknowledges the common criticism of "over-protecting the rights holders beyond the scope of protection provided by traditional copyright laws", and their used model assumed that consumers dislike DRM systems. While definitely informative, it is restricted to one market and was assessing the impact of removal, not how to improve DRM. Another study focusing on undergraduate students' opinion of the topic [9] found that a sizeable amount of their tested population did "not have the basic knowledge about DRM", and while being restricted to university students, this detail would likely reflect the general public.

In terms of "improving" DRM instead of just removing it, there is nothing. There are offered new/modified forms of DRM and/or alternatives that currently exist (see Section V.C), but not how to avoid those same, repeating issues in general for future implementation of current DRM. Nor do any of the studies appear to gather examples of every type, choosing to focus on one or two types or on a specific platform in question.

The only study that comes close to the question posed by this paper was made in 2010 [10] and was focused on the balance between DRM and flexibility for genuine users. While being closer in concept, it still lacks the intent of preventing reoccurring issues, only reducing costs for users.

## 3. METHODOLOGY

In a 2009 medical guideline manual [11], there are five defined steps to the development of evidence-based guidelines:

1. "Identifying and refining the subject area."
2. "Converting and running guideline development groups."
3. "Assessing evidence identified by systematic literature review."

---

4. "Translating evidence into recommendations."
5. "Subjecting the guideline to external review."

While these steps are not entirely compatible with the review provided (see Section IV), the evidence gathered will be used to establish a simple list of advice on how not to approach DRM development. Figure 1 illustrates how the information will be handled throughout this article.

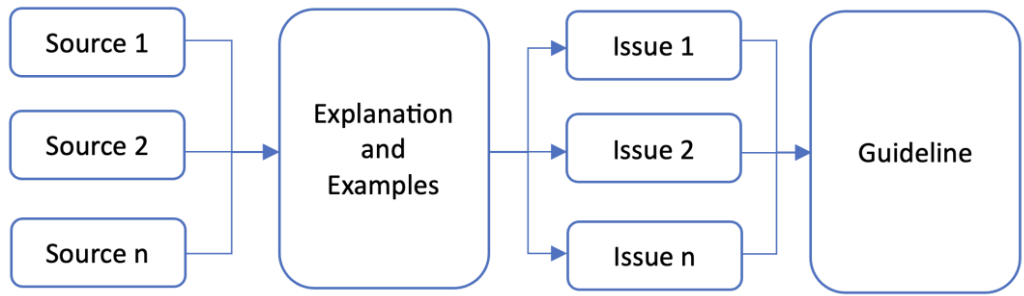

Figure 1. Information handling.

## 4. DRM TECHNOLOGIES AND IMPLEMENTATIONS

### 4.1. Verification

There are three forms of DRM that use verification: Product Keys, Activation Limits, and Always-Online DRM.

### 4.1.1. Product Keys

Product keys are used during installation to ensure the product was purchased genuinely. They are alphanumeric codes ran through an algorithm to determine if it is legitimate. The process of generating product keys is handled via keygens[3]. As DRM, product keys work by an operating system/software package by providing "a card (or email) containing a product key" with its purchase [12]. Without a valid key, no operating system or software.

In the case of Windows, Microsoft specifies its current product key as "a 25-character code that's used to activate Windows and helps verify that Windows hasn't been used on more PCs than the Microsoft Software License Terms allow" [13]. These product keys often come pre-inserted into factory-built computers but are still potentially needed if uninstalling or reinstalling Windows.

### 4.1.2. Activation Limits

Activation limits are a restriction on either the number of active devices a given software can be installed on or the number of installations overall. In terms of DRM, this prevents a user from sharing the software with other non-purchasing users.

Apple is an example of the former. According to Apple Support [14], the limit is "ten devices (no more than five computers) associated with your Apple ID for purchases", in addition to having a ninety-day wait before another Apple ID can be associated with a device once removed.

---

[3] A keygen (key generator) is a program that creates product licensing keys.

### 4.1.3. Always-Online DRM

This form of DRM requires a constant internet-based server connection to access the product as a means of product verification. It is most frequently seen in video games and is considered abhorrent by those who play them.

Denvuo in both of its forms, Anti-Tamper and Anti-Cheat, is a well-known and loathed example of this. Developed by Indeto, it works by "using periodic activation" where any product with Denuvo installed would require the buyers "to log in every few days to authenticate their games", as described by a 2020 study into Video Game DRM [15]. This same paper also demonstrates the biggest possible drawback of this DRM: "Denuvo's servers prevented legitimate buyers from playing Batman Arkham Knight as they couldn't reach the authentication server".

While not explicitly being always-online DRM, the "C-bomb" issue on PlayStation and Xbox consoles can be compared to it. Technically, a form of anti-cheat system, this refers to using the console's built-in CMOS battery[4] and an internet connection to verify an accurate date and time before a game can be started. This primarily refers to digital titles, but in some cases applies to disc-based games as well. However, as IGN India elaborated in 2021 [16], the console will "attempt to sync the date and time over the internet with a remote server" if the CMOS stops working. But if the servers are subsequently unreachable/deactivated, "games and DLC will not work and will be unable to boot". A firmware[5] update released in later September that same year solved the issue for PlayStation consoles but, outside of a statement that they're working on it, there is no sign of it being fixed on Xbox consoles.

## 4.2. Encryption

Encryption is the use of cryptographic[6] keys to obscure a user's data, using two types: symmetric and asymmetric. As Townsend Security [17] defines it, a symmetric key "is used to both encrypt and decrypt the data" while asymmetric keys are "a pair of keys for the encryption and decryption of data", more specifically a public key for encryption and a private key for decryption. This is useful for DRM as it prevents possible crackers[7] from easily seeing the data that they are editing/copying.

However, as a method of DRM, it is not used solely by itself. Its primary usage is as an additional layer of security for other methods. For example, a proposed DRM method formulated by Agarwal, Rana and Pandey in 2018 [18] uses "Dynamic Unidirectional Proxy Re-Encryption and Cipher text Policy Attribute based Encryption techniques", which includes the following phases:

- "authentication check" – which would be used to ensure the user is who they claim to be/that they are allowed to access the product.
- "encryption" – to protect from unauthorised access.
- "data integrity checking" – to ensure no one has tampered with the product.
- "user confirmation" – To again check that the user is who they say they are.
- "data retrieval" – identifying and extracting the data from a database.

---

[4] A CMOS battery is used by a motherboard to retain configuration settings.
[5] Software that allows hardware to function and work with software.
[6] Cryptography refers to methods used to protect data/communication using algorithms.
[7] Someone who removes DRM from software, either to avoid paying for a product or just for fun.

## 4.3. Copy Restriction

Copy restriction, also referred to as anti-piracy, copy protection, copy prevention, content protection, or copy control, is any method used to prevent copyright infringement via the unauthorised copying of media, most commonly video games. It has existed as DRM since software was sold on cassettes and as such, has evolved over the years.

A majority of these methods used by early video games involved either the manual or other such "feelies" that came with the game. The protection in question would activate upon each boot-up and were designed to be fun themselves. As Kotaku reported in 2011 [19], The Secret of Monkey Island's developers used a fondly remembered variant: with the game came a wheel with two sections marked with faces, and the player would have to "rotate the wheel until the corresponding elements matched" the ones provided by the game on boot up and enter the resulting code.

As video games got more complex, so did their copy-restriction methods. Excluding any other DRM that may exist, this is usually carried out by a checksum[8] or a series of checksums that would flag up any alterations to the code. The payloads when this is triggered can be interesting and up to the digression of the product developers.

Some can be quite minor in practice, but massive in impact. Rocksteady employed one such method, as Destructoid reported in 2009 [20]. Batman: Arkham Asylum, one of the company's releases, has a feature that, if it detects the copy being played is pirated, "Batman's cape refused to open…despite pressing the assigned key" making the game impossible after a specific point. It was first discovered due to a thread on Eidos's forums, leading to the admin logging on to explain: "It's not a bug in the game's code, it's a bug in your moral code". A more satirical example can be found in Alan Wake, as IGN reported in a list of similar examples. Its effects were simply "a pretty epic eye-patch" being applied to the main character model and "a gentle reminder to please buy their software in the game's loading screen" [21].

EarthBound was originally released in 1994 in Japan (as Mother 2) and 1995 in the US and, as The Cutting Room Floor [22] states, uses a layered approach. The initial response is a simple region test that throws up an error message. It's after this message is hacked past that the first true layer appears, a subroutine "checks that there's only 8KB of SRAM" since copiers generally have more and produces a warning screen if triggered. Once the game detects that both have been bypassed, the next layer starts affecting the game itself. If a checksum for the previous layer shows a value other than zero, "enemy presence is ramped up to absurd levels" with some areas having "enemies that aren't supposed to be in those locations". In fact, some locations "try to spawn so many enemies that the game outright crashes". After a layer composed of another SRAM[9] checker with an unknown purpose, there is one last checksum. If the game finds that all the previous layers have been disabled, "it hangs and deletes your save files" right before the final boss. Figure 2 represents Earthbound's layered copy protection.

FADE, or as its developer call it DEGRADE, is a copy protection system used in a small number of video games. Most of the games using this system are developed by Bohemia Interactive, with the only known exception being Serious Sam 3: BFE. In an interview with PC Gamer in 2011 [23], the company's CEO elaborated on how it works, with the intention being not to prevent a pirated game from running but "instead (or in addition) to degrade the end user experience of such copies". When activated, DEGRADE can have any effect the developers chose. As stated in that same interview, pirating a game in the ARMA series will lead to "lower accuracy with automatic weapons" and sometimes turning into a bird, while Serious Sam 3: BFE infamously has a "giant invincible armoured scorpion" pursue anyone with an illegitimate copy [24].

---

[8]   A checksum is data from a file that is used to verify it has not been altered.
[9]   Static Random Access Memory

Out of all the possible methods of implementing this form of DRM, Sony BMG used what is possibly the worst one in 2005. They effectively, as a review of Malware and related techniques [25] states, used a "Rootkit to identify and prevent the copying of publications that were made by Sony". It was first exposed by Mark Russinovich on his blog, noting how the software, called Extended Copy Protection or XCP, "uses techniques commonly used by malware to mask its presence" as well as being "poorly written" and having "no means for uninstall" [26]. As a result, multiple anti-virus firms, such as F-Secure started labelling XCP as malware [27].

It took less than two weeks for others to notice the backdoors that XCP left in the systems. Trojans[10] and other backdoor-exploiting malware were being sent around, with hackers even finding an easy way to use it to get around World of Warcraft's anti-cheat program: "only requires that the hacker add the prefix "$sys$" to file names" [28]. Making the situation worse, when Sony BMG issued an uninstaller, it was found by many (including Russinovich) to have its own nearly identical security flaws. Naturally, this all leads to recalls and class-action suits.

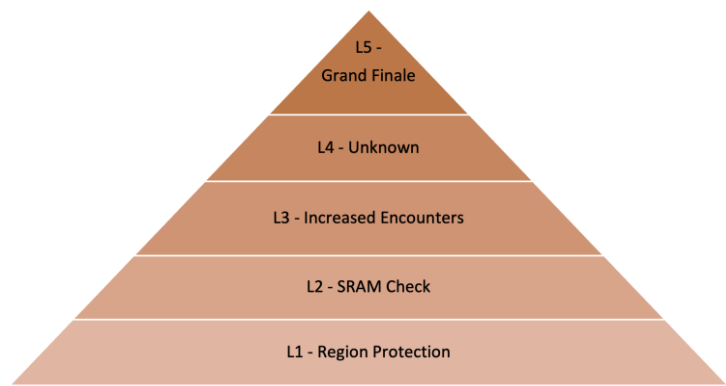

L = Layer

Figure. 2. Pyramid chart representing Earthbound's layered copy protection.

## 4.4. Runtime Restriction

Runtime restrictions refer to the disabling of the software if another software is running that can access the content. These restrictions are used in conjunction with other DRM. While still being copy protection, FADE (discussed in IV.C) is an example of this.

Protected Media Path is an executable designed to create a protected environment, produced by Microsoft for Windows Vista. This environment runs a "separate protected process from the media application" that is designed to "identify untrusted components and plug-ins" so they cannot be loaded [29].

---

[10] A trojan horse is malware that hides its true purpose from a user.

## 4.5. Regional Lockout

Regional lockout is the prevention of hardware/software from running outside of a specific region/territory. There are multiple different ways of performing this, each varying based on what the device/hardware/software is meant for.

One of the most common ways are Disc Regions. Using DVDs as an example, Sony [30] defines this method as "regional coding to prevent the playback of the disc in a geographical area other than the one in which it was released". These codes and their corresponding territories are:

1. Canada, the United States, and other US territories
2. Japan, Europe, South Africa, the Middle East (including Egypt) and Greenland
3. Southeast Asia and East Asia (including Hong Kong)
4. Australia, New Zealand, the Pacific Islands, Central and South America, Mexico, and the Caribbean
5. Eastern Europe, Russia, the Indian Subcontinent, Africa, North Korea, and Mongolia
6. China
7. Unspecified special use
8. International venues such as air and oceanic travel

These correspond with most DVD players sold in these regions, only discs with a matching code will run unless the DVD or the player in question is region free.

For websites, geo-blocking is used to restrict access based on where the user is. Its uses include blocking websites that may be illegal or result in such activities in a specific country, online retailers pushing a user towards their region's version of their website and streaming services offering different libraries customised for each region. As Avast [31] explained, if geo-blocking detects that the user's "IP address is connected to a region or country where a particular site's content is blocked", the user will be unable to access that content. There are multiple ways to bypass geo-blocking such as VPNs[11] and proxy servers which hide a user's IP address and/or allow them to modify where it appears to be coming from.

## 4.6. Tracking

Tracking allows software/document owners to ensure that their software/documents "are only being used by authorised people" and in authorised locations [32]. There are two forms of Tracking: Watermarking and Metadata.

### 4.6.1. Watermarks

A watermark, as defined by Adobe [33], is "a logo, piece of text or signature superimposed onto a photograph". It should be noted that videos can have watermarks, but image watermarks are the more commonly seen used version. A study into watermarking in 2018 lists several attacks that may be carried out on watermarks, ranging from trying to "remove the watermark or simply make it undetectable" to inserting "a new valid watermark", or even just "breaking the security method in watermarking techniques" [34].

### 4.6.2. Metadata

---

[11] Virtual Private Network

Metadata is, as Avast [35] defines it, "the hidden data the accompanies every image, video and file" in a device. It provides various pieces of information about the image/video/file such as: the time and data it was created, author, file size, and location made, among others. Exactly what this entails can vary based on device type; for example, location metadata from a smartphone will use coordinates while a laptop would not.

As DRM, a potential implementation [36] would be "controlled in real-time after deployment, with access and function attributes able to changed" while still being "consistent with commercial deployment and procurement processes". This would be a more active form of DRM. However, as of the time of writing, there have been no attempts at such a system.

## 4.7. Hardware

Finally, there are several possible forms of hardware-based DRM. Regional Lockout (see IV.E for more information) is one such example.

Another can be found in Flexplay, which was a short-lived alternative to DVD rentals, lasting from 2003 to 2009. As The Museum of Obsolete Media [37] states, the intent was "a means for rental of new films without the need for returning discs after use" via "a mechanism to make the disc unplayable after a certain length of time". These discs were vacuum-sealed, and "the bonding resin holding the inner and outer layers together reacts to oxygen", which would render the disc unplayable within 48 hours of opening and, after a set amount of time, unopened. However, a lack of interest and environmental concerns led to the format being abandoned.

## 5. CURRENT LAWS, OPPOSITIONS AND ALTERNATIVES

### 5.1. Laws

Strictly speaking, DRM falls under various copyright laws, with some countries having different laws/ approaches regarding it.

Internationally, DRM circumvention laws are enacted as a requirement of the World Intellectual Property Organisation Copyright Treaty. The WIPO Copyright Treaty was established in 1996 and works in conjunction with the Berne Convention, which was established in 1889 and "deals with the protection of works and the rights of their authors" [38]. In relation to the Treaty, this means that "any Contracting Party (even if it is not bound by the Berne Convention) must comply with the substantive provisions" it defines. Additionally, it expands the recognised author rights to include "the right of distribution", "the right of rental", and "a broader right of communication to the public" [39].

However, some countries are not signatories and/or have their own copyright laws.

In China, the WIPO Copyright Treaty is carried out via its Interim Regulations. However, a study in 2018 about DRM in China [40] found that there were four key areas of challenge: "conflict between DRM and copyright law, legal issues related to DRM and management, limited features of DRM, and Chinese law supporting anti-circumvention". India, however, was not a signatory of WIPO until 2013, but was one of the Berne Convention before then. Their Copyright Act was established in 1957, and as an overview in 2022 [41] states, provides "an economic right to the author", "a paternity right", "an integrity right", and "a general right". Israel is not a signatory of WIPO and has no law against circumventing DRM or similar measures.

Countries in the European Union use a different implementation of the WIPO Copyright Treaty: The Copyright and Information Society Directive 2001. The intent of this directive was to contribute "to the achievement of these objectives" via "harmonisation of the laws of the Member States on copyright and related rights" [42]. Out of the Member States at the time, only Greece and Denmark met the deadline for implementation, with four others implementing in 2003 and the remaining eight being referred to the European Court of Justice. In 2019, it was revised into the Directive on Copyright in the Digital Single Market. There are several national implementations of the Copyright and Information Society Directive. In the UK, the Directive was transposed into the Copyright and Related Rights Regulations 2003 [43].

The Digital Millennium Copyright Act is the U.S.' WIPO Copyright Treaty implementation and was established in 1998. According to the U.S. Copyright Office [44], it is an amendment that focused on:

1. "establishing protections for online service providers in certain situations if their users engage in copyright infringement"
2. "encouraging copyright owners to give greater access to their works in digital formats by providing them with legal protections against unauthorised access to their works"
3. "making it unlawful to provide false copyright management information…or to remove or alter that type of information in certain circumstances".

The Copyright Claims Board was established in 2020 after the Copyright Alternative in Small-Claims Enforcement Act was passed the same year, and is an alternative method to "resolve copyright disputes of a relatively low economic value" [45].

## 5.2. Oppositions and Criticism

Tracking allows software/documents owners to ensure that their software/documents "are only being used by authorised people" and in authorised locations [32]. There are two forms of Tracking: Watermarking and Metadata.

Strictly speaking, DRM falls under various copyright laws, with some countries having different laws/ approaches regarding it.

There are many organisations and individuals that oppose the use of DRM for various reasons. The most frequently used is that DRM's goal of being uncrackable is a futile goal at best, and a problem-causing restriction at worst.

Defective by Design [46] is an online initiative that began in 2006 by the Free Software Foundation, and is dedicated to "eliminate DRM as a threat to innovation in media, the privacy of readers, and freedom for computer users". Personal privacy in digital media, for instance, which is a very complex topic, has also been investigated by other researchers [47]. To this end, the "defective by design" has done several campaigns to this effect, the most well-known being their yearly "Day Against DRM". In 2013, they protested Netflix's plan to implement "streaming videos in HTML5, only in browsers with "Premium Video Extensions" [48]. This would involve Encrypted Media Extensions, a standard for web videos that often attracted criticism due to "no safeguards whatsoever for accessibility, security research or competition" [49]. In spite of the campaign's efforts (and the outside controversy), the World Wide Web Consortium greenlit EME for use in 2017.

CD Projekt S.A. is a popular game developer, with their "Red" department known for their The Witcher series and, in a less positive fashion, Cyberpunk 2077. Two other things they are well-known for are the digital distribution platform GOG (for more information, see section V.C) and, as of 2012, releasing games DRM-free. In an interview with Forbes in 2012 [50], the company's CEO Marcin Iwiński states that DRM "does not work" and that it is "cracked within hours of the release of every single game", rendering the implementation "wasted money and development".

## 5.3. Alternatives

Naturally, alternatives both legal and illegal, exist for media with DRM. While not strictly being an alternative, GOG is a digital distribution platform that does not use any form of DRM, stating that they put "gamers first and respect their need to own games" [51]. Being made by CD Projekt S.A. (see section V.B for more information), every game sold on their website, regardless of if they have any involvement in its production, is completely free of DRM even if it has DRM in its other releases.

Piracy is the most common form of DRM alternatives, mostly due to the fact that any DRM has been manually removed from the product. However, between the possible punishments if caught and the very high chance of malware accompanying the software, piracy is often also quite inconvenient.

One proposed but not implemented alternative is the Artistic Freedom Voucher. Proposed in 2003 by Dean Baker [52], the Artistic Freedom Voucher is "an alternative mechanism for supporting creative and artistic work". The idea is that creative workers and/or their intermediaries "register with the government in the same way that religious or charitable organisations" do and place the work in the public domain, after which they'd receive "a certain amount of money" as a "refundable tax credit".

# 6. GUIDELINE PROPOSAL

## 6.1. Issues to Address

As it stands, there are multiple issues with DRM as it currently is. Any proposal for guidelines will need to address each one to a relevant and appropriate degree. The following are the most prominent issues; others of a lesser degree will not be discussed here.

### 6.1.1. Trust

There are two ways in which trust must be addressed with DRM: public trust in it, and the developers' trust in its protection.

In terms of the public, they have had several reasons to not trust DRM. Events such as the Sony BMG scandal (Section IV.C) have eroded public confidence in DRM, with Denuvo's reputation (Section IV.A.3) not being much better. Any set of guidelines would ideally need to address most if not all prior issues in this regard.

### 6.1.2. Malicious Potential

This would need to be addressed in a manner that deals with every possibility, even those that haven't been done/used yet. These would include backdoors/rootkit-like behaviour (for more information, see the Sony BMG scandal in Section IV.C), botnets, other malware types/malware-like behaviour, bricking a device (rendering it unfunctional due to firmware/hardware damage).

### 6.1.3. Preservation

While the desire to protect the rights of a product are to be respected, it should not be at the expense of genuine consumers, no matter how old the product is. The guidelines will attempt to address this fine line. For more information of examples of this issue, see the "C-bomb" in Section IV.A.3 and Flexplay in Section IV.G.

### 6.1.4. The Guideline

1. No form of Digital Rights Management should attempt, intentionally or otherwise, to damage and/or destroy either firmware or hardware as a result of activation.
2. No form of Digital Rights Management should mimic malware in its operation.
3. No form of Digital Rights Management should attempt to create, intentionally or otherwise, backdoors or holes in a device's security.
4. No form of Digital Rights Management should monitor any user's device, regardless of activation.
5. No form of Digital Rights Management should compromise the performance of a device regardless of activation.
6. No form of Digital Rights Management should remain on a device after the software/hardware it came with is disabled/removed/uninstalled.
7. In the event that the Digital Rights Management does linger or need to otherwise be disabled independently of the software/hardware it came with, the uninstallation program should be subject to these same standards.
8. Copy Protection should not impact a legitimate user's experience.
9. Always-Online-DRM should only be attempted if the developers have the appropriate number of servers/servers with the appropriate level of strength.
10. Activation limits should have a remote deactivation feature for disconnecting devices without needing the device in question.
11. Runtime restrictions should not be set to trigger to software needed for device operation.
12. No hardware-based Digital Rights Management should deactivate a legitimate user's device over time.
13. If it is discovered that this may occur with a device after the method has been implemented, a way to disable/prevent this should be established and applied at no additional cost to the user.
14. No product that uses hardware-based Digital Rights Management should degrade for a legitimate user.

## 7. DISCUSSION

The main contribution of this study is proposing a guideline that helps organisations implement DRM. The study results and the framework used in this research can also be used by researchers to develop DRM production guidelines for specific digital assets based on countries, governments, and organisations' requirements.

Overall, the answer to the question posed (see Section I) is quite simple yet also complex to answer.

For the first part of the question ("What negative implementations of DRM exist?") that was simple. Throughout the main literature review, both positive and negative examples were found. Primarily in Sections IV.A.3, IV.C and IV.G, negative implementations can be seen to have massive effects but seemingly little in terms of consequences. Outside of Sony BGM causing lawsuits, the only real negative effect for the DRM, is the reduction of trust in the security measure. So much so that, as can be seen in the initial literature review (Sections II and IV), there is little to no other option considered other than "get rid of it".

As for "what would need to be done to prevent them from reoccurring?", that may vary. On the surface, it is quite simple as well. The guidelines above do address the issues displayed specifically in addition to some potential issues. It is regarding their usefulness that it gets complex. As these are merely guidelines, there is little actually making developers/those who would wish to implement DRM follow them.

Additionally, there is the issue of limitations. There is little to no documentation from companies/vendors regarding the nature of the DRM used outside of that which is used to promote it. If any current guidelines exist within these organisations, they are not publicly available. As a result, this article cannot take them into account. Also, there is no way to read the code of the used DRM, so this article does not consider how it is coded, only the effects of the finalised product.

These guidelines are far from definitive. While encompassing the gathered negative examples, there is the possibility that something has been overlooked, that the guidelines themselves are too vague or restrictive, or that a new form of DRM/brand new example occurs after this article.

## 8. CONCLUSION

DRM is a rarely acknowledged factor in this digital age. At least, by the correct parties. While not many of the current implementations feature major problems, the ones that do seem to get away with relatively minor consequences. Even with two (or three, depending on how one looks at it) instances of rootkit-like behaviour by DRM, there are no standards for their creation.
While this guideline could address this if implemented, ultimately, it would be entirely up to the developers themselves if they are followed.

Overall, the most pressing factor in play, both generally and with the guideline specifically, is trust. Would the public regain trust in DRM if these guidelines or similar ones were implemented? Can developers be trusted to take the guidelines to heart if there is little making them? Can developers trust that consumers will still buy a product with DRM?

## AUTHORS

**SHANNON KATHLEEN COATES** received Upper Second Class Honours for Cyber Security at Anglia Ruskin University, Cambridge. Her research interests include cyber security, malware analysis, and cyber security incident prevention.

**HOSSEIN ABROSHAN** is a Senior Lecturer at Anglia Ruskin University, Cambridge. He received a PhD degree in Cyber Security (business economics) from Ghent University. He also has over 25 years of experience in the IT and information security fields in the financial, telecom, maritime, manufacturing, and research sectors. He holds several professional certifications, including Certified Information Security Manager (CISM) and ISO 27001 Lead Auditor. His research interests include social engineering and psychological aspects of cybersecurity, security of critical infrastructures, and AI for cyber security.